\documentclass[RNAAS]{aastex62}

\graphicspath{{./}{figures/}}

\begin{document}

\title{Is PSR J0250+5854 at the Hall attractor stage?}

\correspondingauthor{Andrei P. Igoshev}
\email{ignotur@gmail.com}

\author[0000-0003-2145-1022]{Andrei P. Igoshev}
\affiliation{Department of Physics, Technion, Haifa 3200003, Israel}

\author[0000-0002-4292-8638]{Sergei B. Popov}
\affiliation{Lomonosov Moscow State University, Sternberg Astronomical Institute\\
Universitetski pr. 13, Moscow 119234 Russia}
\affiliation{National Research University ``Higher School of Economics'', Department of Physics\\
Myasnitskaya str. 20, Moscow 101000, Russia}


\begin{abstract}
In this note we propose that recently discovered radio pulsar J0250+5854 with 23.5 sec spin period is presently at the Hall attractor stage.
This can explain low temperature and absence of magnetar-like activity of this source together with its spin period and period derivative.
We present results of calculations of the evolution of this source in a simple model of magnetic  field decay. The neutron star could start its evolution as a magnetar with initial field $\sim 10^{14}-10^{15}$~G for realistic range of parameter $Q$ responsible for crust imperfections.
Future measurements of surface temperature and age of this neutron star might help to probe this hypothesis.
\end{abstract}

\section{} 

Recently, \cite{tan2018} presented a remarkable discovery of a radio pulsar --- PSR J0250+5854, ---  with spin period 23.5 sec. Its magnetic field estimated with the magneto-dipole formula is $2.6 \times 10^{13}$~G. The authors also presented upper limits on X-ray emission of the pulsar which exclude surface temperature above $\sim 85$~--$110$~eV. It was suggested that the pulsar has a non-magnetar origin. In this research note we propose a different scenario for formation of PSR J0250+5854.

This neutron star (NS), as already noted by \cite{tan2018}, looks similar in many respects to seven near-by isolated NSs detected due to their surface thermal radiation (so-called {\it Magnificent seven}, M7). In \cite{popov2010} three types of NSs --- magnetars, M7, and radio pulsars --- were described in a unified manner in the framework of decaying magnetic field. In our opinion, PSR J0250+5854 can also fit this picture (so, it is on the evolutionary track from magnetars to M7 and further on) with one addition to the evolutionary scenario --- the Hall attractor. Previous numerical modeling demonstrates that as soon as this stage is reached, rapid magnetic field evolution governed by Hall cascade ceases (see \citealt{2016PNAS..113.3944G} and references therein).

During last few years the idea of the Hall attractor is actively discussed in application to NSs with no significant magnetic field evolution in the core (see initiation of this discussion in \citealt{gc2014}). Moderately evolved NSs with initially large fields at ages $\sim 1$~Myr can be at this stage. Studies of X-ray properties of RX J1856.5-3754 which is one of the M7 sources, did not show features predicted for the Hall attractor \citep{ptt2017}. Thus, M7 can still be at a preceding stage. Here we propose that PSR J0250+5854 can be on this stage.

Our estimates presented below are based on the model described in \cite{ip2018}. We assume that the magnetic field evolves due to a Hall cascade and Ohmic dissipation. The timescale of the Hall cascade at a given moment is inversely proportional to magnetic field of the NS. Ohmic resistivity is controlled by phonons till the NS is hot enough, and by crust imperfections afterwards. Thermal evolution (without direct URCA processes) is followed in a simplified manner based on fits of two models of cooling valid for magnetars \citep{vigano} and normal NSs \citep{shternin}. Period evolution is calculated using the magneto-dipole formula, alignment is neglected.

The Hall attractor stage is switched on after three initial characteristic timescales of the field decay \citep{gc2014}. Since this moment, magnetic energy dissipation is small, and so there is no additional heating of a NS and magnetar-like activity is absent.  We search for initial magnetic field $B_0$ of PSR J0250+5854 fixing the parameter $Q$ responsible for crust imperfections such a way to get observed values of  period $P$ and period derivative $\dot P$. 
NSs with different initial $B_0$ and $Q$ reach observed values of $P$ and $\dot P$ at different ages, so their red-shifted NS surface temperature is a function of $Q$ and formation path. Results are presented in the Figure.

We see that for a wide (and realistic) ranges of $B_0$ and $Q$ it is possible to obtain parameters equal to those of PSR J0250+5854. For all cases with the Hall attractor presented in the Figure the NS has reached the Hall attractor stage. Thus, our assumption that PSR J0250+5854 was born as a magnetar and now reached a quiet period due to saturation of the Hall cascade agrees with observations. 

Up to now no NSs are identified as objects at the Hall attractor stage. To clarify the case of PSR J0250+5854 it is necessary to obtain an independent estimate of its age (maybe, via study of kinematics of this source), and to detect and study properties of the surface thermal emission. 
If X-ray/UV luminosity of PSR J0250+5854 is below $L_{UV} = 3\times 10^{31}$~erg~s$^{-1}$ (corresponding to the surface temperature below 40~eV) the Hall attractor stage provides the most plausible explanation of its rotational and thermal properties.

Identification of a NS with known $P, B$, age, and surface temperature at the Hall attractor stage would help to probe its interior, in particular to estimate the parameter $Q$. Understanding of the Hall attractor is important not only for studies of evolution of isolated NS, but also for so-called accreting magnetars in X-ray binaries \citep{ip2018}.

\begin{figure}
\begin{center}
\includegraphics[scale=0.85,angle=0]{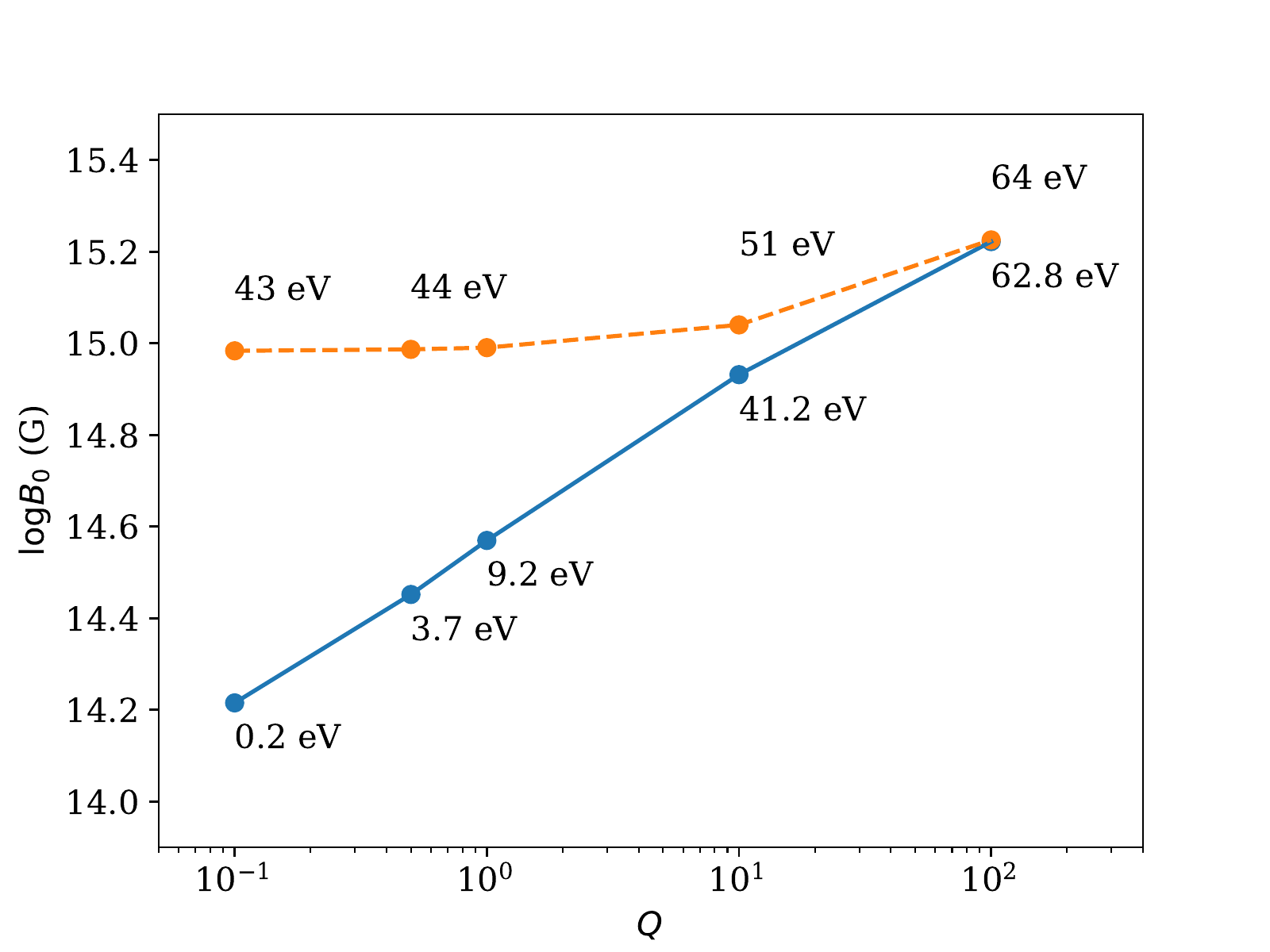}
\caption{Initial magnetic field $B_0$ vs. imperfection parameter $Q$. The line corresponds to combinations of $B_0$ and $Q$ for which present-day parameters --- $P$ and $\dot P$, --- of PSR J0250+5854 are reproduced and the NS is at the Hall attractor stage (solid line), or the Hall cascade continues (dashed line). Text labels are the red-shifted surface temperatures. \label{f:qb0}}
\end{center}
\end{figure}

\acknowledgments
SP acknowledges support from the RSF grant 14-12-00146.

\bibliographystyle{aasjournal} 
\bibliography{bib}

\end{document}